\documentclass[ reprint,nofootinbib,amsmath,amssymb,aps]{revtex4-1}
\usepackage{graphicx}
\usepackage{subfigure}
\usepackage{dcolumn}
\usepackage[colorlinks,linkcolor=magenta,anchorcolor=cyan,citecolor=blue]{hyperref}
\usepackage{bm}
\usepackage[multiple]{footmisc}

\def\be{\begin{equation}}
\def\ee{\end{equation}}
\def\ba{\begin{eqnarray}}
\def\ea{\end{eqnarray}}

   \def\y {\psi}    \def\a {\alpha}                  
        \def\F {\Phi}      \def\.{\cdot}

\begin{document}
\title{Strong Cosmic Censorship in Einstein-Maxwell-Scalar-Gauss-Bonnet Theory}
\author{Aofei Sang}
\email{aofeisang@mail.bnu.edu.cn}
\author{Jie Jiang}
\email{Corresponding author. jiejiang@mail.bnu.edu.cn}
\affiliation{College of Education for the Future, Beijing Normal University at
Zhuhai, 519087, China}
\affiliation{Department of Physics, Beijing Normal University, Beijing 100875, China\label{addr2}}

\begin{abstract}
Recently, it is found that the strong cosmic censorship (SCC) is violated in RNdS black hole by a minimal coupled neutral massless scalar field at the linear perturbation level. For the Einstein-Maxwell-Scalar-Gauss-Bonnet theory which is famous for the spontaneous scalarization phenomenon, there exists the scalar field coupled with the Gauss-Bonnet term. Whether the SCC of the RNdS black hole can be repaired by the non-minimal coupled scalar field perturbation at the linear level becomes an interesting question. In this paper, we firstly investigate the extendibility of the metric beyond the Cauchy horizon in the Einstein-Maxwell-Scalar-Gauss-Bonnet theory. Then, we examine the SCC of the RNdS black hole by calculating the quasinormal mode of the non-minimal coupled scalar field and find that the SCC can be recovered under some parameters.
\end{abstract}
\maketitle
\section{Introduction}
It is widely believed that the future of our world can be completely determined by a series of initial values. In general relativity (GR), this means that the physics of the entire spacetime can be determined by the Einstein equation and initial values scattered across a Cauchy surface. However, it is proved by Hawking and Penrose that the singularity, where we cannot describe by the metric, is ubiquitous in our spacetime \cite{HP1970,Penrose:1969pc}. And, the appearance of the time-like singularity implies the breakdown of the determinism. To avoid this situation, Penrose proposed the strong cosmic censorship (SCC) conjecture \cite{Penrose:1969pc}, which states that the singularity must be space-like or null, or equivalently, the initial value should be inextendible at the Cauchy horizon. A modern version of SCC proposed by Christodoulou gives a more precise statement \cite{Christodoulou:2008nj}: the spacetime metric cannot be extended beyond the Cauchy horizon in the form of weak solutions to the field equations.

For the asymptotic case, such as the Kerr black hole and Reissner-Nordstrom black hole, the SCC is confirmed to be out of a problem. It is demonstrated that a linear perturbation outside the black hole with inverse power decay will be amplified due to the blueshift effect \cite{Price:1972pw,Dafermos:2010hb,Dafermos:2014cua}.
This amplification turns the Cauchy horizon into a mass inflation singularity eventually such that the metric is inextendible beyond the Cauchy horizon \cite{Poisson:1990eh,Dafermos:2012np,Dafermos:2003wr,Luk:2017jxq,Luk:2017ofx}. But for the black hole in de Sitter spacetime, the positive cosmological constant will cause an exponential decay of the perturbation matter field, which might be able to compete with the blueshift effect \cite{Brady:1996za,Costa:2014aia,Hintz:2015jkj}. Recently, the SCC is found to be violated by a massless neutral scalar field in Reissner-Nordstrom de Sitter (RNdS) black hole in the nearly extremal region \cite{Cardoso:2017soq}, which arises lots of attention. Later, other violations are also found on the RNdS background by the massless charged scalar field, the massless Dirac field, etc. \cite{Mo:2018nnu,Ge:2018vjq,Liu:2019rbq,Dias:2018ufh}. Moreover, the validity of the SCC for the RNdS black hole considering the non-minimal coupled scalar field perturbation in some modified gravitational theory is also investigated \cite{Destounis:2019omd}.

Many astronomical observations and recent gravitational wave observations show that although GR is in good agreement with the observed phenomena in the weak-field region, it still needs to be corrected in high curvature or high energy regions \cite{LIGOScientific:2016lio,Yunes:2016jcc,LIGOScientific:2018dkp,LIGOScientific:2020tif,EventHorizonTelescope:2021dqv}. The Einstein-Scalar-Gauss-Bonnet (ESGB) theory draw lots of attention since a new phenomenon called spontaneous black hole scalarization will occur in this theory \cite{Silva:2017uqg,Doneva:2017bvd,East:2021bqk}. In many black hole solutions of ESGB theory, we can obtain stable non-trivial scalar field hairs, which will cause the observational results that are different from the GR case in the strong-field region but still agree with the current observations in the weak-field region \cite{Silva:2017uqg,Doneva:2017bvd,East:2021bqk,Antoniou:2017acq}. In ESGB theory, the scalar field is coupled with the Gauss-Bonnet term. Then, it is natural to ask that whether the SCC of the RNdS black hole can be repaired by these non-minimal coupled scalar field perturbation at the linear level. Therefore, in this paper, we would like to examine that whether the decay rate of the non-minimal coupled scalar field is slow enough to make the Cauchy horizon a singularity in the RNdS black hole.

Our paper is organized as follows. In Sec. \ref{sec2}, we give the equation of motion of the Einstein-Maxwell-Scalar-Gauss-Bonnet (EMSGB) theory and introduce the background spacetime. We also give the radial equation of motion of the scalar field through variable separation and obtain the asymptotic solution. In Sec. \ref{sec3}, we study the extendibility of the metric beyond the Cauchy horizon and find the condition such that the SCC is respected. In Sec. \ref{sec4}, we review the numerical methods used to calculate the QNM. We also show some relevant results in this section. We draw a conclusion in Sec. \ref{sec5}.

\section{Einstein-Maxwell-Scalar-Gauss-Bonnet Theory}\label{sec2}
In this paper, we consider the EMSGB theory, in which the massless scalar field $\Phi$ nonminimally couples to a Gauss-Bonnet term. The full action of this theory is given by \cite{Antoniou:2017acq}
\be\begin{aligned}\label{action}
S=&\frac{1}{16\pi}\int d^4 x\sqrt{-g}\left[R-2\Lambda-F_{ab}F^{ab}\right.\\
&\left.-\frac{1}{2}\nabla_a \Phi\nabla^a \Phi+f(\Phi)R_{\text{GB}}^2\right]\,,
\end{aligned}\ee
where $R$ is the Ricci scalar, $\Lambda$ is the positive cosmological constant, $F_{ab}=\nabla_a A_b-\nabla_b A_a$ is the electromagnetic tensor, and $R_{\text{GB}}^2=R_{abcd}R^{abcd}-4R_{ab}R^{ab}+R^2$ is the quadratic Gauss-Bonnet term which is coupled with the scalar field by a coupling function $f(\Phi)$. In this paper, we would like to consider a special case in which
\ba
f(\F)=\frac{1}{2}\a \F^2
\ea
with the coupling constant $\a$. The equations of motion can be obtained by the variation of Eq. \eqref{action}, and are given by \cite{Antoniou:2017acq}
\be\begin{aligned}\label{eomall}
&G_{ab}+\Lambda g_{ab}=T_{ab}^{\text{sc}}+T_{ab}^{\text{EM}}\,,\\
&\nabla^2\Phi+\alpha \Phi R^2_{\text{GB}}=0\,,\\
&\nabla^a F_{ab}=0\,,
\end{aligned}\ee
respectively. Here, we use
\be\begin{aligned}\label{Tsc}
T_{ab}^{\text{sc}}=&-\frac{1}{4}g_{ab}\nabla_c\Phi\nabla^c\Phi+\frac{1}{2}\nabla_a\Phi\nabla_b\Phi\\
&-\frac{\alpha}{4}\left(g_{ca}g_{db}+g_{da}g_{cb}\right)\eta^{edfg}\tilde{R}^{ch}{}_{fg}\nabla_h\nabla_e \Phi^2
\end{aligned}\ee
with
\be\begin{aligned}
 \tilde{R}^{ab}{}_{fg}=\eta^{abcd}R_{cdfg}=\frac{\epsilon^{abcd}}{\sqrt{-g}}R_{cdfg}
\end{aligned}\ee
to denote the energy-momentum tensor of the scalar field part and
\be\begin{aligned}\label{TEM}
T_{ab}^{\text{EM}}=2F_a{}^{c}F_{bc}-\frac{1}{2}F_{cd}F^{cd}g_{ab}
\end{aligned}\ee
to denote the energy-momentum tensor of the electromagnetic field part.

As is considered in the previous papers, we would like to consider a charged spherical symmetric solution to the field equation, i.e., RNdS metric, as the background. The line element of this spacetime is given by
\be\begin{aligned}\label{metric}
ds^2=-f(r)dt^2+\frac{1}{f(r)}dr^2+r^2 d\Omega^2\,,
\end{aligned}\ee
where
\be\begin{aligned}\label{blackening1}
f(r)=1-\frac{2M}{r}+\frac{Q^2}{r^2}-\frac{\Lambda r^2}{3}\,.
\end{aligned}\ee
This is a special solution of the EMSGB theory with vanishing scalar field. 

If we assume that $r_c$, $r_+$ and $r_-$ are the cosmological horizon, event horizon and Cauchy horizon respectively. Then, the blackening factor can be rewritten as
\be\begin{aligned}\label{blackening2}
f(r)=\frac{\Lambda}{3r^2}(r_c-r)(r-r_+)(r-r_-)(r-r_o)\,,
\end{aligned}\ee
where $r_o$ is the minimum root of $f(r)=0$ and we can find $r_o=-r_c-r_+-r_-$ through the fact that Eq. \eqref{blackening1} is same with Eq. \eqref{blackening2}. Furthermore, we can define the surface gravity of each horizon as
\be\begin{aligned}
\kappa_i=\frac{1}{2}|f'(r_i)|\quad\text{with}\quad i=\{c,+,-,o\}\,.
\end{aligned}\ee

Then, we view the coupled scalar field $\Phi(t,r,\theta,\phi)$ as a perturbation on this RNdS background spacetime. Considering the symmetries of the spacetime, the scalar field can be expanded to \cite{Konoplya:2011qq}
\be\begin{aligned}\label{sepvar}
\Phi(t,r,\theta,\phi)=\sum_{lm}e^{-i \omega t}Y_{lm}(\theta,\phi)\frac{\psi(r)}{r}\,,
\end{aligned}\ee
in which $Y_{lm}(\theta,\phi)$ is the spherical harmonics. With Eqs. \eqref{metric} and \eqref{sepvar}, the equation of motion \eqref{eomall} satisfied by the scalar field can be simplified to an one dimention Schr$\ddot{\text{o}}$dinger-like equation
\be\begin{aligned}\label{schr}
\frac{d^2 \psi(r)}{dr^2_{\ast}}+[\omega^2-V(r)] \psi(r)=0
\end{aligned}\ee
with
\be\begin{aligned}\label{potential}
V(r)=&\frac{f(r)}{r^2}\left[l(l+1)+r f'(r)\right.\\
&\left.-4\alpha(f'(r)^2-f''(r)+f(r)f''(r))\right]\,.
\end{aligned}\ee
Note that we used the tortoise coordinate
\be\begin{aligned}
dr_{\ast}=\frac{dr}{f(r)}
\end{aligned}\ee
in Eq. \eqref{schr}. If we consider the physical region between $r_+$ and $r_c$, we can obtain
\be\begin{aligned}\label{rs}
r_{\ast}=&-\frac{1}{2\kappa_c} \ln\left(1-\frac{r}{r_c}\right)+\frac{1}{2\kappa_+}\ln\left(\frac{r}{r_+}-1\right)\\
&-\frac{1}{2\kappa_-}\ln\left(\frac{r}{r_-}-1\right)+\frac{1}{2\kappa_o}\ln\left(1-\frac{r}{r_o}\right)\,.
\end{aligned}\ee
From Eq. \eqref{potential}, we can also notice that the effective potential vanishes on every horizon. Then, it is easy to find that the asymptotic solution of Eq. \eqref{schr} near each horizon is
\be\begin{aligned}\label{asymptotic}
\psi\sim e^{\pm i\omega r_{\ast}}\,,\quad r\rightarrow r_{c,+,-,o}\,,
\end{aligned}\ee
in which $e^{i \omega r_{\ast}}$ represents the outgoing wave and the other one represents the ingoing wave. With some physical limitations \cite{Konoplya:2011qq}, we require that there is only ingoing wave near the event horizon $r_+$ and only outgoing wave near the cosmological horizon $r_c$, i.e.,
\be\begin{aligned}\label{boundarycon}
&\psi\sim e^{-i\omega r_{\ast}}\,,\quad r\rightarrow r_+\,,\\
&\psi\sim e^{i\omega r_{\ast}}\,,\quad\,\,\,\, r\rightarrow r_c\,.
\end{aligned}\ee
The solutions of Eq. \eqref{schr} satisfied the above boundary condition \eqref{boundarycon} are called quasinormal mode (QNM). It is clear that the frequencies $\omega$ of QNM in these solutions are discrete.

\section{Strong cosmic censorship and quasinormal mode}\label{sec3}

In this section, we would like to study the relationship between SCC and QNM in EMSGB theory.

SCC requires that the metric cannot extend beyond the Cauchy horizon, otherwise, there will be a region where the physics cannot be determined by the initial values. Thus, to investigate the SCC, we need to study the extendibility of the solution to the equation of motion beyond the Cauchy horizon in EMSGB theory. Following the argument in Ref. \cite{Destounis:2019omd}, if the solution is extendible beyond the Cauchy horizon, in other words, if there exists a weak solution at the Cauchy horizon, the integral
\be\begin{aligned}\label{extend1}
\int_{\mathcal{V}}d^4x\sqrt{-g}(G_{ab}+\Lambda g_{ab}-T_{ab}^{\text{sc}}-T_{ab}^{\text{EM}})\Psi
\end{aligned}\ee
must be vanishing. Here, $\Psi$ is a smooth, compactly supported test function and we use $\mathcal{V}$ to denote the integral domain, which is a small neighborhood.

As in the general RNdS case, the finiteness of the first two terms requires the Christoffel symbol is locally square-integrable in $\mathcal{V}$. The last term is finite since the electric potential $A_a$ is regular near the Cauchy horizon. Finally, we will examine the integrability of the third term
\be\begin{aligned}\label{extend2}
\int_{\mathcal{V}}d^4x\sqrt{-g}T_{ab}^{\text{sc}}\,.
\end{aligned}\ee
According to Eq. \eqref{Tsc}, there are three terms in the integral. The integral of the first two terms is the same as the case of the general minimum coupled scalar field. It can be schematically denoted as
\be\begin{aligned}\label{extend3}
\int_{\mathcal{V}}d^4x\sqrt{-g}(\partial\Phi)^2\Psi\,.
\end{aligned}\ee
The finiteness of Eq. \eqref{extend3} leads to $\Phi\in H^1_{\text{loc}}$, which means the first-order derivative is locally square-integrable. Similarly, the third term of Eq. \eqref{extend2} can be expressed as
\be\begin{aligned}\label{extend4}
\int_{\mathcal{V}}d^4x\sqrt{-g}R(\partial^2\Phi^2)\Psi
\end{aligned}\ee
schematically. With a similar discussion of Ref. \cite{Destounis:2019omd}, we can find
\ba\begin{aligned}\label{extend5}
&\int_{\mathcal{V}}d^4x\sqrt{-g}R(\partial^2\Phi^2)\Psi\\
&\lesssim \sup (\Psi)\int_{\mathcal{V}}d^4x\sqrt{-g}\left[(\partial^2\Phi^2)^2+R^2\right]\,.
\end{aligned}\ea
To ensure that the integral is finite, we demand $\Phi^2\in H^2_{\text{loc}}$, i.e., $\Phi^2$ is square-integrable up to the second-order derivative.

Then, with the asymptotic solution \eqref{asymptotic} obtained in Sec. \ref{sec2}, it is easily to find
\be\begin{aligned}
\Phi\sim &A_1 e^{-i\omega (t-r_{\ast})}+A_2e^{-i\omega (t+r_{\ast})}\\
=&A_1 e^{-i\omega u}+A_2e^{-i\omega u}e^{-2i\omega r_{\ast}}
\end{aligned}\ee
near the Cauchy horizon $r_-$, where $A_1$, $A_2$ are the superposition coefficients and  $u=t-r_{\ast}$. Since $u$ is regular near $r_-$, $e^{-i\omega u}$ does not contribute to the infiniteness of the integral \eqref{extend5}. Therefore, only $e^{-2i\omega r_{\ast}}$ matters when we consider the regularity of Eq. \eqref{extend5} near $r_-$. Considering that the dominant part of $r_{\ast}$ is
\be\begin{aligned}
r_{\ast}\sim -\frac{1}{2\kappa_-}\ln\left|\frac{r}{r_-}-1\right|
\end{aligned}\ee
when $r$ approaches to $r_-$, the only part that we should consider in $\Phi$ reduces to
\be\begin{aligned}\label{extendphi}
\Phi\sim e^{i\frac{\omega}{\kappa_-}\ln|r-r_-|}=|r-r_-|^{i \frac{\omega}{\kappa_-}}\,.
\end{aligned}\ee
The frequencies can be written as $\omega=\omega_R+i \omega_I$, where $\omega_R$ and $\omega_I$ represent the real part and the imaginary part of $\omega$ respectively. Then, since $|r-r_-|^{i \omega_R/\kappa_-}$ is oscillatory, we can find that $|r-r_-|^{\beta}$ is the term that ultimately determines the integrability, where we defined
\be\begin{aligned}
\beta\equiv - \omega_I/\kappa_-\,.
\end{aligned}\ee

According to the above analysis, the extendibility of the metric beyond the Cauchy horizon requires $\Phi^2\in H^2_{\text{loc}}$. Using Eq. \eqref{extendphi}, this requirement means that
\be\begin{aligned}
\int_{\mathcal{V}}dr(\partial_r^2\Phi^2)^2\sim&\int_{\mathcal{V}}dr |r-r_-|^{4(\beta-1)}\\
=&\frac{1}{4\beta-3}|r-r_-|^{4\beta-3}
\end{aligned}\ee
is finite near $r_-$, i.e.,
\be\begin{aligned}
\beta > \frac{3}{4}\,.
\end{aligned}\ee
Then, since the SCC requires that the metric is inextendible beyond the Cauchy horizon, it will be respected if there exists a mode such that $0<\beta\leq 3/4$ with nonvanishing coupling constant $\a$ . Note that we only care about the positive $\beta$, since the negative $\beta$ leads to superradiation.

\section{Numerical methods and results}\label{sec4}

In this section, we would like to calculate the QNM numerically. In the first part, we will introduce the numerical methods. In the second part, we will show some relevant results.

\subsection{Numerical methods}
There are lots of numerical methods to calculate the accurate QNM frequencies \cite{Konoplya:2011qq}. In this paper, we use the pseudospectral method \cite{Jansen:2017oag,Miguel:2020uln} to calculate the QNM and use the direct integration method \cite{Chandrasekhar:1975zza,Molina:2010fb} to verify the correctness of the results. We also use the six-order WKB approximation \cite{Konoplya:2003ii} to calculate the QNM at the large-$l$ limit.

The pseudospectral method translates the continuous differential equation to a discrete metric equation by expanding the wave function by a cardinal function formed by the Chebyshev polynomial.

First, we divide the interval $[-1,1]$ into $N$ small intervals using $N+1$ grid points. In general, we choose
\be\begin{aligned}
x_i=\cos\left(\frac{i}{N}\pi\right)\,,\quad i=0,1,\dots,N
\end{aligned}\ee
as the grid points, which are called Chebyshev grid. Then, a regular function $F$ can be approximated as
\be\begin{aligned}\label{speF}
F(x)=\sum^N_{i=0}F(x_i)C_i(x)\,,
\end{aligned}\ee
where $C_i(x)$ is the cardinal function which satisfied $C_i(x_j)=\delta_{ij}$. This property ensures the value of $F(x)$ is accuracy on the grid points. With Eq. \eqref{speF}, it is easy to find the first and second order derivatives of $F$ can be written as
\be\begin{aligned}\label{speD}
&F'(x)=
%&\sum^N_{i=0}F(x_i)C'_i(x)=\sum^N_{i=0}\sum^N_{j=0}F(x_i)C'_i(x_j)C_j(x)\\
\sum^N_{i=0}\sum^N_{j=0} F(x_i)D^{(1)}_{ij}C_j(x)\,,\\
&F''(x)=
%&\sum^N_{i=0}F(x_i)C''_i(x)=\sum^N_{i=0}\sum^N_{j=0}F(x_i)C''_i(x_j)C_j(x)\\
\sum^N_{i=0}\sum^N_{j=0} F(x_i)D^{(2)}_{ij}C_j(x)
\end{aligned}\ee
with the derivative matrix
\be\begin{aligned}
D^{(1)}_{ij}=C'_i(x_j)\quad \text{and} \quad D^{(2)}_{ij}=C''_i(x_j)\,.
\end{aligned}\ee
For the Chebyshev grid we chose, the cardinal function can be expressed as the linear combination of the Chebyshev polynomial $T_n(x)$:
\be\begin{aligned}\label{speC}
&C_i(x)=\frac{2}{Np_i}\sum^N_{m=0}\frac{1}{p_m}T_m(x_i) T_m(x)\,,\\
&p_0=p_N=2\,,\quad p_j=1\,,\quad\quad (j=1,2\dots,N-1)\,.
\end{aligned}\ee
With this expression, we can easily calculate the derivative matrix \eqref{speD}. Moreover, it should be noted that Eq. \eqref{speF} can not be used for approximating the function with divergence and oscillation since the cardinal function formed by the Chebyshev polynomial is smooth over the entire interval. Therefore, when we use the Chebyshev polynomial expansion, we should firstly rescale the wave function such that the function is regular including the boundary. Besides, according to the grid, we also need to rescale the coordinate to $[-1,1]$.

Second, we would like to consider a second order linear ordinary differential equation (ODE) with a general form:
\ba\begin{aligned}\label{speeqy}
a_0(\omega,x)F(x)+a_1(\omega,x)F'(x)+a_2(\omega,x)F''(x)=0\,.\quad\quad
\end{aligned}\ea
Here, we assume the function $F(x)$ is already satisfies the requirement we mentioned above. Since Eq. \eqref{speeqy} should establish at each grid point, Together with Eqs. \eqref{speF} and \eqref{speD}, the ODE becomes
\be\begin{aligned}\label{speeqylisan}
&\sum^N_{j=0}\left[a_0(\omega,x_i)\delta_{ji}+a_1(\omega,x_i)D^{(1)}_{ji}+a_2(\omega,x_i)D^{(2)}_{ji}\right]y(x_j)\\
&=0
\end{aligned}\ee
for any $i=0,1,\dots,N$. If the highest power of $\omega$ is two in the coefficients, we can expand the coefficients as $a_{i}(\omega,x)=a_{i0}(x)+\omega a_{i1}(x)+\omega^2 a_{i2}(x)$ with $i=0,1,2$, where the second index of $a$ is used to denote the power of $\omega$. Then, Eq. \eqref{speeqylisan} can be further written as
\be\begin{aligned}\label{speeqymatrix}
\left(\tilde{M}_0+\omega\tilde{M}_1+\omega^2\tilde{M}_2\right)\tilde{Y}=0\,,
\end{aligned}\ee
in which
\be\begin{aligned}
\tilde{Y}=\left(y(x_0)\,,y(x_1)\,,\dots\,,y(x_N)\right)^T
\end{aligned}\ee
and
\be\begin{aligned}
(\tilde{M}_{\mu})_{ij}=a_{0\mu}(x_i)\delta_{ji}+a_{1\mu}(x_i)D^{(1)}_{ji}+a_{2\mu}(x_i)D^{(2)}_{ji}
\end{aligned}\ee
with $\mu=0,1,2$. Note that every $\tilde{M}_{\mu}$ can be calculated numerically now. We can further simplify \eqref{speeqymatrix} by defining
\be Y=\begin{pmatrix}
\tilde{Y}\\
\omega\tilde{Y}
\end{pmatrix}\ee
and
\be
M_0=
\begin{pmatrix}
\tilde{M}_0 & \tilde{M}_1\\
0 & \mathbb{I}
\end{pmatrix}
\,,\quad
M_1=
\begin{pmatrix}
0 & \tilde{M}_2\\
-\mathbb{I} & 0
\end{pmatrix}
\,.
\ee
Then, Eq. \eqref{speeqymatrix} is equivalent to
\be\begin{aligned}\label{speeigen}
(M_0+\omega M_1) Y=0\,.
\end{aligned}\ee
In Eq. \eqref{speeigen}, $M_0$ and $M_1$ are all numerical matrices, thus, to obtain the QNM frequencies is to calculate the eigenvalue of $(-M_1^{-1}M_0)$ and $y(x)$ can be obtained by the corresponding eigenvector through Eq. \eqref{speF}. It can be easily found that there will be $2(N+1)$ eigenvalues if we take $N+1$ grid points. However, some of them might be fake. To avoid the inauthentic modes, we need to take different $N$ to calculate the eigenvalues and pick out the stable modes. Moreover, we can find there exists a nontrivial solution of $Y$ if and only if $\det(M_0+\omega M_1)=0$, which gives another way to find $\omega$.

Finally, before using the pseudospectral method to calculate Eq. \eqref{schr}, we can notice that the boundaries of $\psi(r)$ are $r_+$ and $r_c$ and $\psi(r)$ is highly oscillating near the two boundary from Eq. \eqref{boundarycon}. Therefore, we need to rescale Eq. \eqref{schr} first. To eliminate the oscillation, we can extract the oscillatory parts in $\psi(r)$. The oscillatory parts near $r_+$ and $r_c$ are
\ba\begin{aligned}
\y(r)\propto \left(\frac{r-r_+}{r_+}\right)^{\frac{i \omega}{2\kappa_+}}\,,\quad\quad r\to r_+\,,\\
\y(r)\propto \left(\frac{r_c-r}{r_c}\right)^{-\frac{i\omega}{2\kappa_c}}\,,\quad\quad r\to r_c\,,\\
\end{aligned}\ea 
individually. Then, let
\be\begin{aligned}\label{respsi}
\psi(r)=\left(\frac{r-r_+}{r_+}\right)^{\frac{i \omega}{2\kappa_+}}\left(\frac{r_c-r}{r_c}\right)^{-\frac{i\omega}{2\kappa_c}}\tilde{\psi}(x)
\end{aligned}\ee
with
\ba
r=\frac{r_c-r_+}{2}x+\frac{r_c+r_+}{2}\,.
\ea
After substiuting Eq. \eqref{respsi} into Eq. \eqref{schr} and multiplying both side of Eq. \eqref{schr} by
\be\begin{aligned}\label{restilde}
\frac{1}{(r-r_+)(r-r_c)}\left(\frac{r-r_+}{r_+}\right)^{\frac{i \omega}{2\kappa_+}}\left(\frac{r_c-r}{r_c}\right)^{-\frac{i\omega}{2\kappa_c}}\,,
\end{aligned}\ee
we get a equation about $\tilde{\psi}(x)$ which is regular and the definitional domain is $[-1,1]$. We would not show the rescaled equation since it is interminable. When $\alpha=0$, our modes fit perfectly with the previous data for the RNdS case, which verifies the correctness of our result partially.

To ensure that our result is reliable, we need to use various methods to compute. In this paper, we use the direct integration method to check the data of the pseudospectral method. First, near the boundary, we can solve Eq. \eqref{schr} by expanding $\psi(r)$ as
\be\begin{aligned}\label{zjjfexp}
&\psi(r)=(r-r_+)^{\frac{i\omega}{2\kappa_+}}\sum_{n=0}^\infty\psi_n^+(r-r_+)^n\,,\quad r\rightarrow r_+\,,\\
&\psi(r)=(r-r_c)^{-\frac{i\omega}{2\kappa_c}}\sum_{n=0}^\infty\psi_n^c(r-r_c)^n\,,\quad r\rightarrow r_c\,.
\end{aligned}\ee
Replacing $\psi(r)$ in Eq. \eqref{schr} by Eq. \eqref{zjjfexp}, we can obtain the series coefficients $\psi_n^+$ and $\psi_n^c$. Then, if we assume $\omega=\omega_0$ is a specific complex number, using Eq. \eqref{zjjfexp} and the coefficients we obtained, we can get a pure numerical result of both $\psi(r)$ and $\psi'(r)$ near $r_+$ as the boundary condition. With this boundary condition, we can solve Eq. \eqref{schr} in the interval $(r_+, (r_++r_c)/2)$ using \emph{Mathematica}. Furthermore, we can also solve Eq. \eqref{schr} in $((r_++r_c)/2, r_c)$ with the boundary condition near $r_c$ by the same process. Combining the two solutions together, we can get a piecewise function in $(r_+, r_c)$. Then, we can obtain the acceptable $\omega_0$ by requiring the first order derivative of the piecewise function is continuous at $(r_++r_c)/2$.

\begin{figure*}
    \centering
    \subfigure[$\beta=-{\omega_I}/{\kappa_-}$ for $\Lambda=0.06$, $\alpha=0.02$]{
    \includegraphics[width=0.46\textwidth]{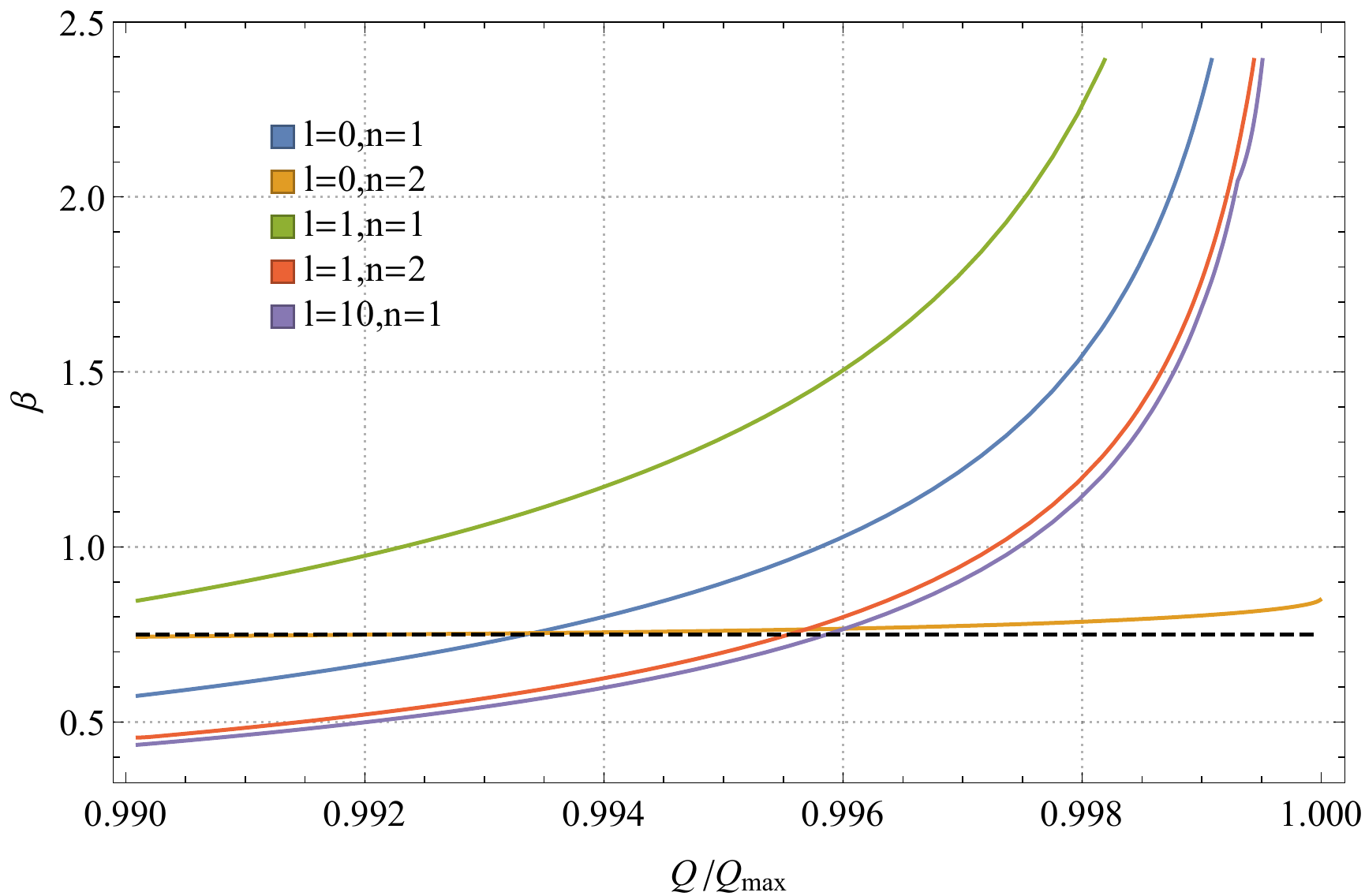}
    %\caption{fig1}
    \label{fig1a}
    }
    \subfigure[$\beta=-{\omega_I}/{\kappa_-}$ for $\Lambda=0.06$, $\alpha=0.12$]{
    \includegraphics[width=0.46\textwidth]{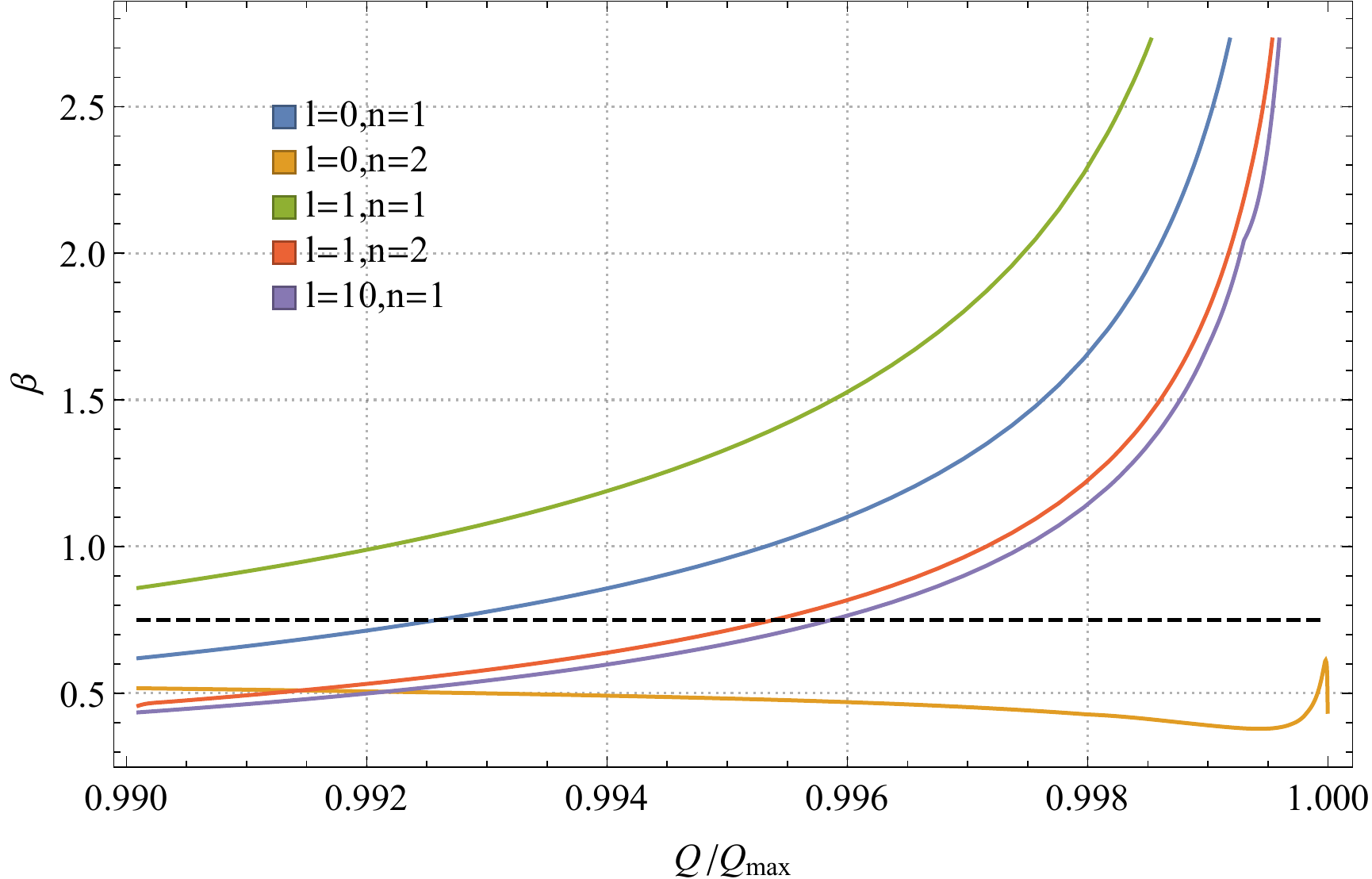}
    \label{fig1b}
    }
    \quad
    \subfigure[$\beta=-{\omega_I}/{\kappa_-}$ for $\Lambda=0.14$, $\alpha=0.02$]{
    \includegraphics[width=0.46\textwidth]{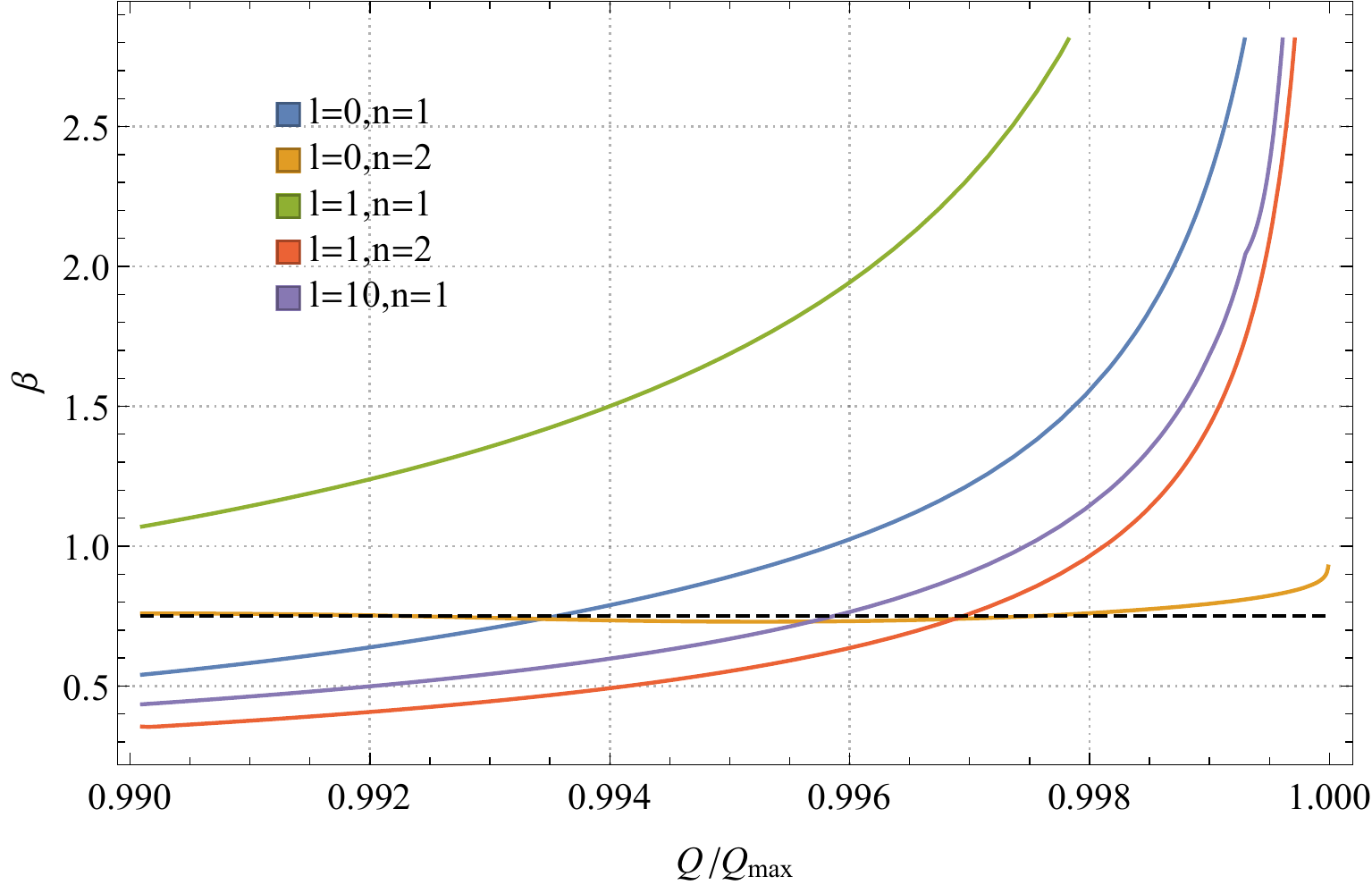}
    \label{fig1c}
    }
    \subfigure[$\beta=-{\omega_I}/{\kappa_-}$ for $\Lambda=0.14$, $\alpha=0.12$]{
    \includegraphics[width=0.46\textwidth]{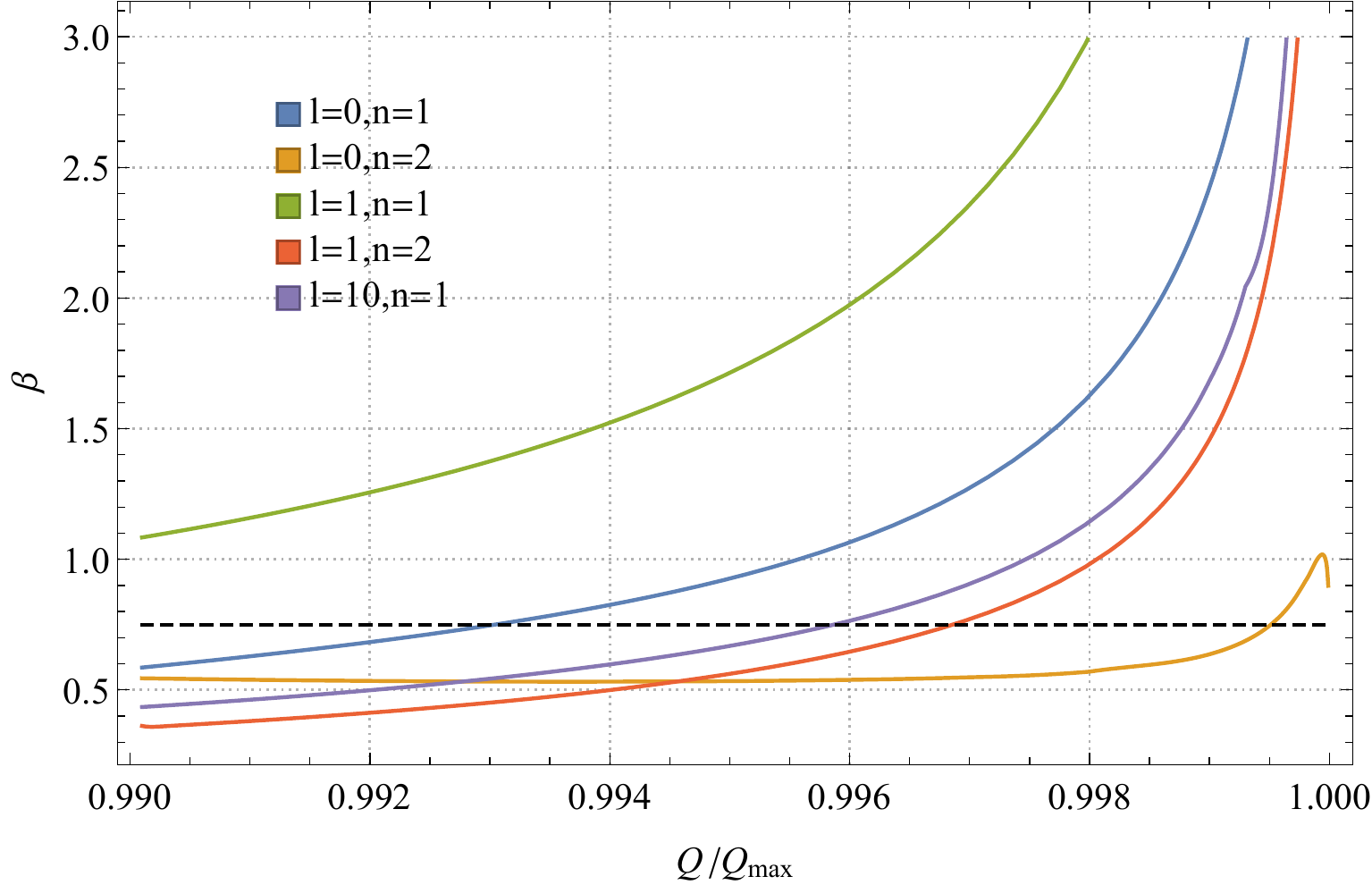}
    \label{fig1d}
    }
    \caption{Dominant and sub-dominant NE modes ($l=0$), dS modes ($l=1$) and the dominant PS mode ($l=10$), divided by $\kappa_-$, for different cosmological constant and coupling constant. We use $n$ to denote the overtone numbers. The black dashed line is for $\beta=3/4$}
    \label{fig1}
\end{figure*}

\begin{table*}[ht]
    \centering
    \renewcommand\arraystretch{1.9}
    \begin{tabular}{|c|c|c|c|c|}
    \hline
    %\diagbox{a}{$\frac{\omega}{\kappa_-}$}{b} %添加斜线表头
    & $l=0$ & $l=1$ & $l=2$ & $l=10$ \\
    \hline
    $\alpha=0$ & $0.00002692-0.00003973 i$ & $0.00007847-0.00002947 i$ & $0.00013534-0.00002812 i$ & $0.0005787-0.00002749 i$ \\
    \hline
    $\alpha=0.05$ & $0.000028828-0.00003924 i$ & $0.00007924-0.00002943 i$ & $0.0001357-0.000028109 i$ & $0.0005788-0.00002749 i$\\
    \hline
    $\alpha=0.1$ & $0.00003062-0.00003879 i$ & $0.000080002-0.00002940 i$ & $0.00013616-0.00002809 i$ & $0.0005789-0.00002749 i$ \\
    \hline
    $\alpha=0.15$ & $0.00003235-0.000038398 i$ & $0.00008075-0.00002937 i$ & $0.0001366-0.000028074 i$ & $0.00057901-0.00002749 i$\\
    \hline
    $\alpha=0.2$ & $0.00003399-0.00003804 i$ & $0.000081503-0.000029348 i$ & $0.00013699-0.00002805 i$ & $0.0005791-0.00002749 i$ \\
    \hline
    \end{tabular}
    %\caption{这是一张三线表}\label{tab:aStrangeTable}  标题放在这里也是可以的
    % 这个地方是引入注释或者标注
%\begin{tablenotes}
 %   \item[1] A represent B
  %  \item[2] B represent C
%\end{tablenotes}
    \caption{The dominant QNMs ${\omega}/{\kappa_-}$ for $\Lambda =0.06$ and $Q=0.2$.}\label{tab1}%添加标题 设置标签
\end{table*}

\subsection{Relevant results}

In this subsection, we would like to show some representative results. According to the previous paper, the SCC is hardly violated for the RNdS black holes that are not near extremal. The situation is similar to the EMSGB theory. In Table \ref{tab1}, we show some results of $\omega/\kappa_-$ for $\Lambda =0.06$, $Q=0.2$ and we will not show the modes with positive imaginary part since it leads to the superradiation, which is irrelevant with SCC. We take $\alpha=0,\,0.05,\,0.1,\,0.15,\,0.2$. Moreover, considering the coupling term should be a small correction of the Einstein-Maxwell theory, we will only give the data for small coupling constant $\alpha$ from $0$ to $0.2$ in this paper. For each $\alpha$, we calculate the dominant mode, i.e., the mode with largest imaginary part, for different multipole numbers: $l=0$, $l=1$, $l=2$, $l=10$. We can find that all the absolute values of the imaginary part in Table \ref{tab1} are far less than $3/4$, i.e., $\beta\ll 3/4$. There is no violation of SCC on this occasion. Therefore, we will concentrate on the nearly extremal black holes in the following discussion.

Next, we can also divide the QNMs to three families: near-extremal (NE) modes ($l=0$), de Sitter (dS) modes ($l=1$), photon sphere (PS) modes ($l=10$). In Fig. \ref{fig1}, we show the pictures of how the dominant and sub-dominant NE modes, dS modes, and the dominant PS mode of the non-minimal coupled scalar field evolved on the nearly extremal RNdS black hole varies with the black hole charge. Fig. \ref{fig1a} lies on the top left corner in Fig. \ref{fig1} is for $\Lambda=0.06$ and $\alpha=0.02$, while the top right one Fig. \ref{fig1b} is for the same $\Lambda$ and $\alpha=0.12$. Fig. \ref{fig1c} lies on the bottom left corner is for $\Lambda=0.14$ and $\alpha=0.02$, while the top right one is for a same $\Lambda$ and $\alpha=0.12$. The black dashed line is $\beta=3/4$, which is the criteria to determine whether the SCC is violated. According to Fig. \ref{fig1}, we can find that for a nearly extremal black hole, the dominant QNM will be made up of two different modes. And when a black hole gets more and more extreme, the dominant mode will eventually depend on the dominant NE mode.

\begin{figure}
\centering
\includegraphics[width=0.48\textwidth]{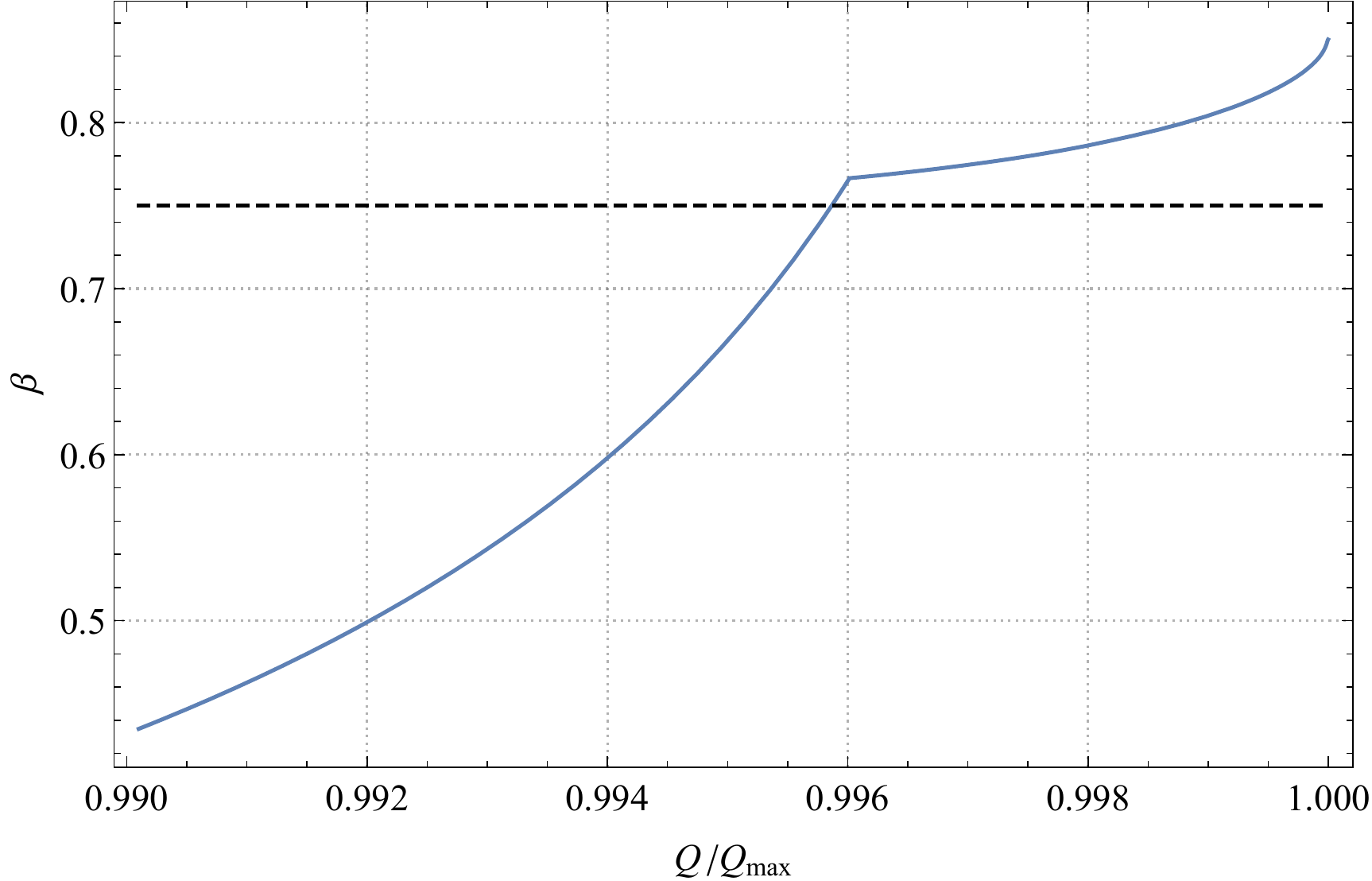}
\caption{Dominant modes $\beta=-\omega_I/\kappa_-$ for $\Lambda=0.06$ and $\alpha=0.02$.}
\label{fig21}
\end{figure}

\begin{figure}
\centering
\includegraphics[width=0.48\textwidth]{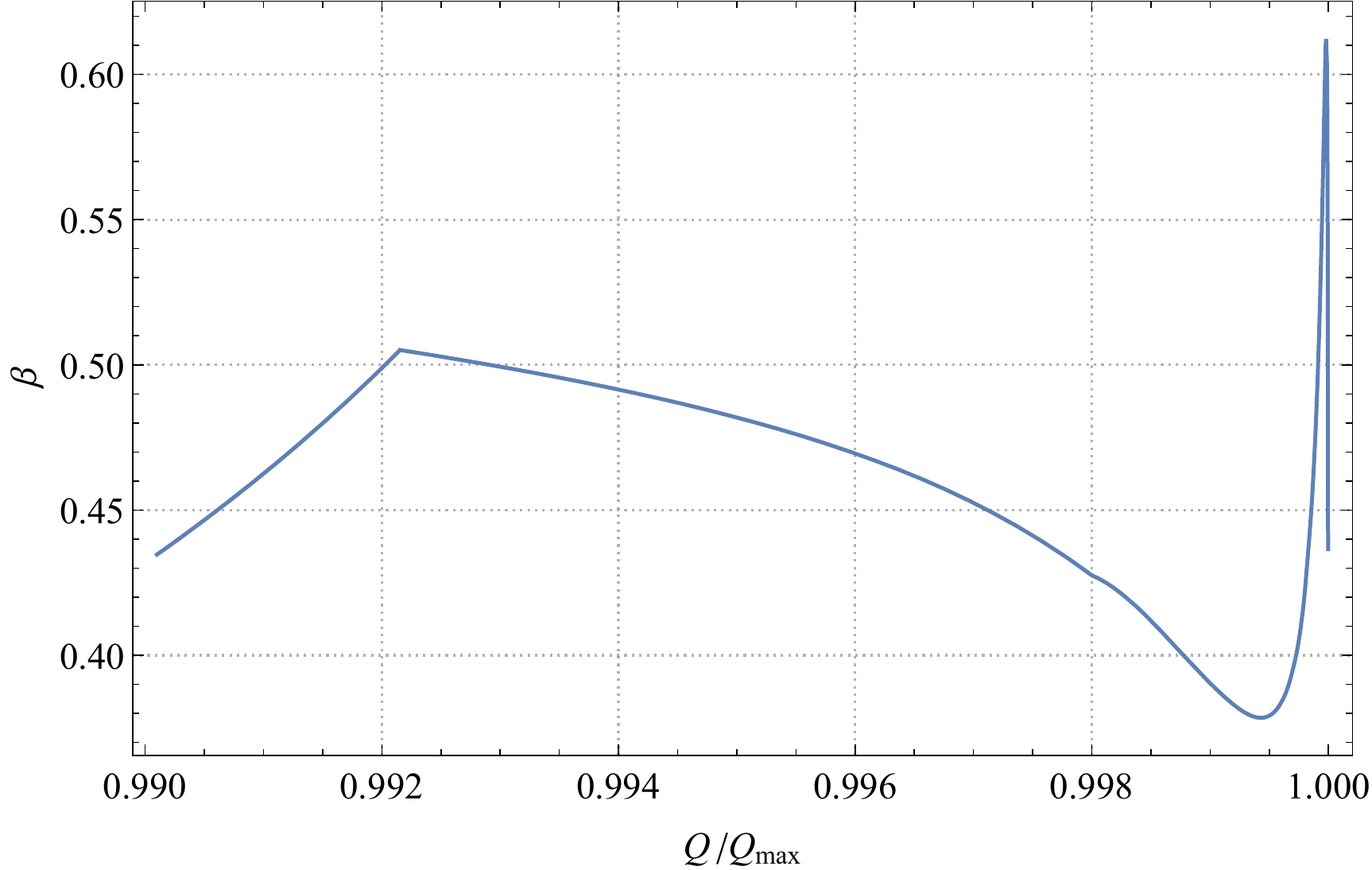}
\caption{Dominant modes $\beta=-\omega_I/\kappa_-$ for $\Lambda=0.06$ and $\alpha=0.12$.}
\label{fig22}
\end{figure}

Then, we can extract the dominant modes through these modes. We present the dominant modes for $\Lambda=0.06$, $\alpha=0.02$ in Fig. \ref{fig21} and the dominant modes for $\Lambda=0.06$ and $\alpha=0.12$ in Fig. \ref{fig22} as examples. The dominant modes of these two figures are both composed of the PS modes and the NE modes. We can easily find that the dominant modes are larger than $3/4$ for some charge parameter in the case of $\Lambda=0.06$ and $\alpha=0.02$, therefore, SCC is violated in the theory with $\Lambda=0.06$ and $\alpha=0.02$. While for $\Lambda=0.06$, $\alpha=0.12$, the dominant modes are always below the black dashed line, which means the SCC will be respected in this case.

\begin{figure}
\begin{minipage}[b]{.8\linewidth}
\centering
\includegraphics[width=2.8in,height=1.79in]{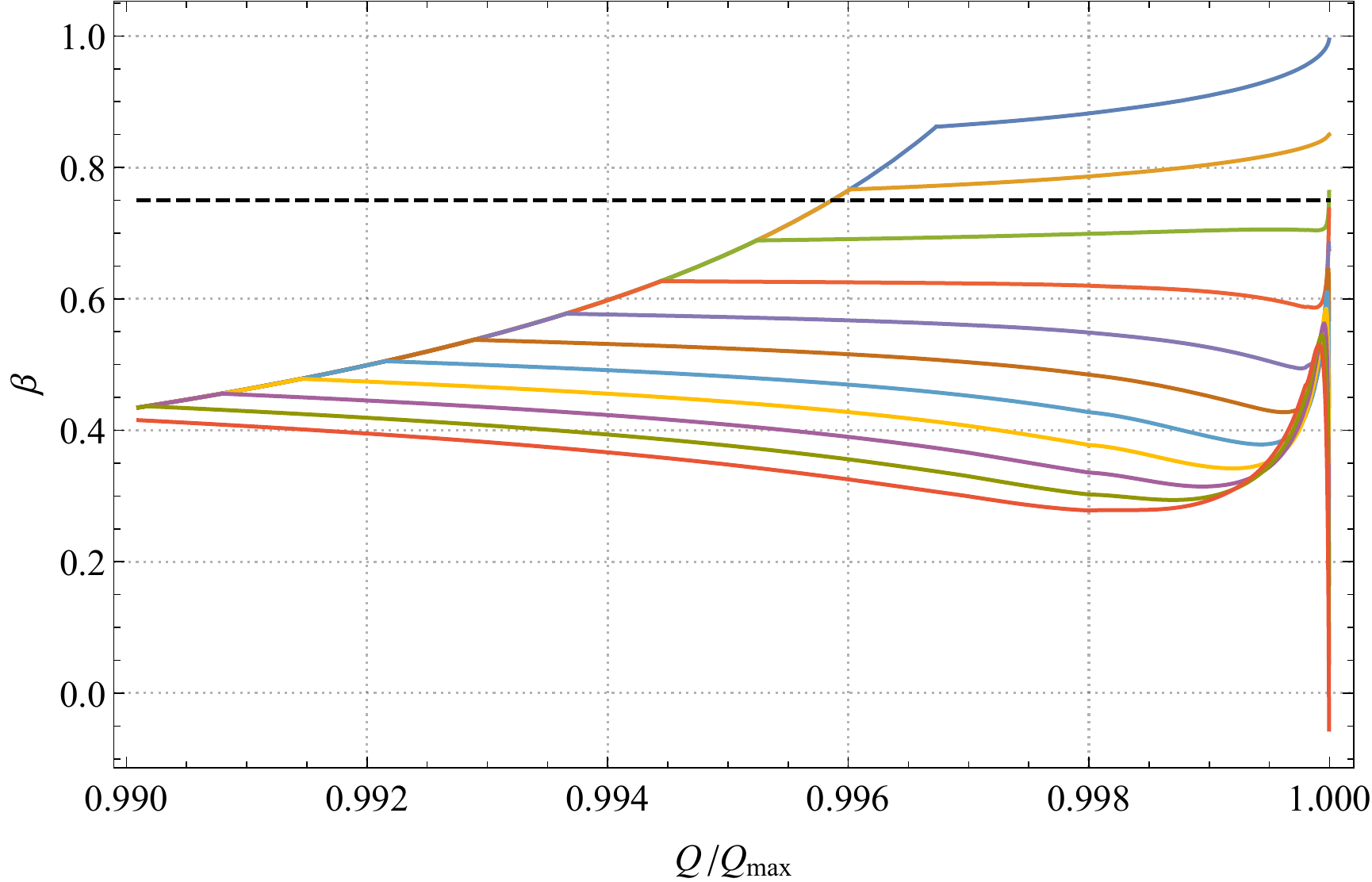}\label{a}
\end{minipage}
\begin{minipage}[b]{.18\linewidth}
\centering
\includegraphics[width=0.3in,height=1.4in]{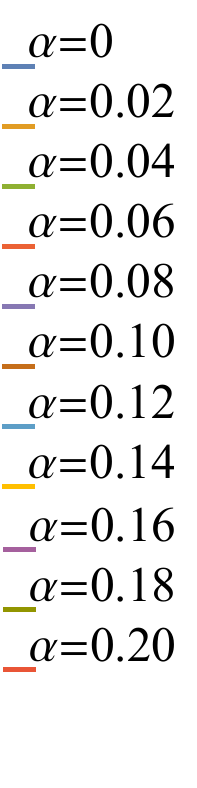}
\label{b}
\end{minipage}
\caption{Dominant modes $\beta=-\omega_I/\kappa_-$ for $\Lambda=0.06$ and $\alpha$ from $0$ to $0.2$, spaced $0.02$ apart.}
\label{fig31}
\end{figure}

\begin{figure}
\begin{minipage}[b]{.8\linewidth}
\centering
\includegraphics[width=2.8in,height=1.79in]{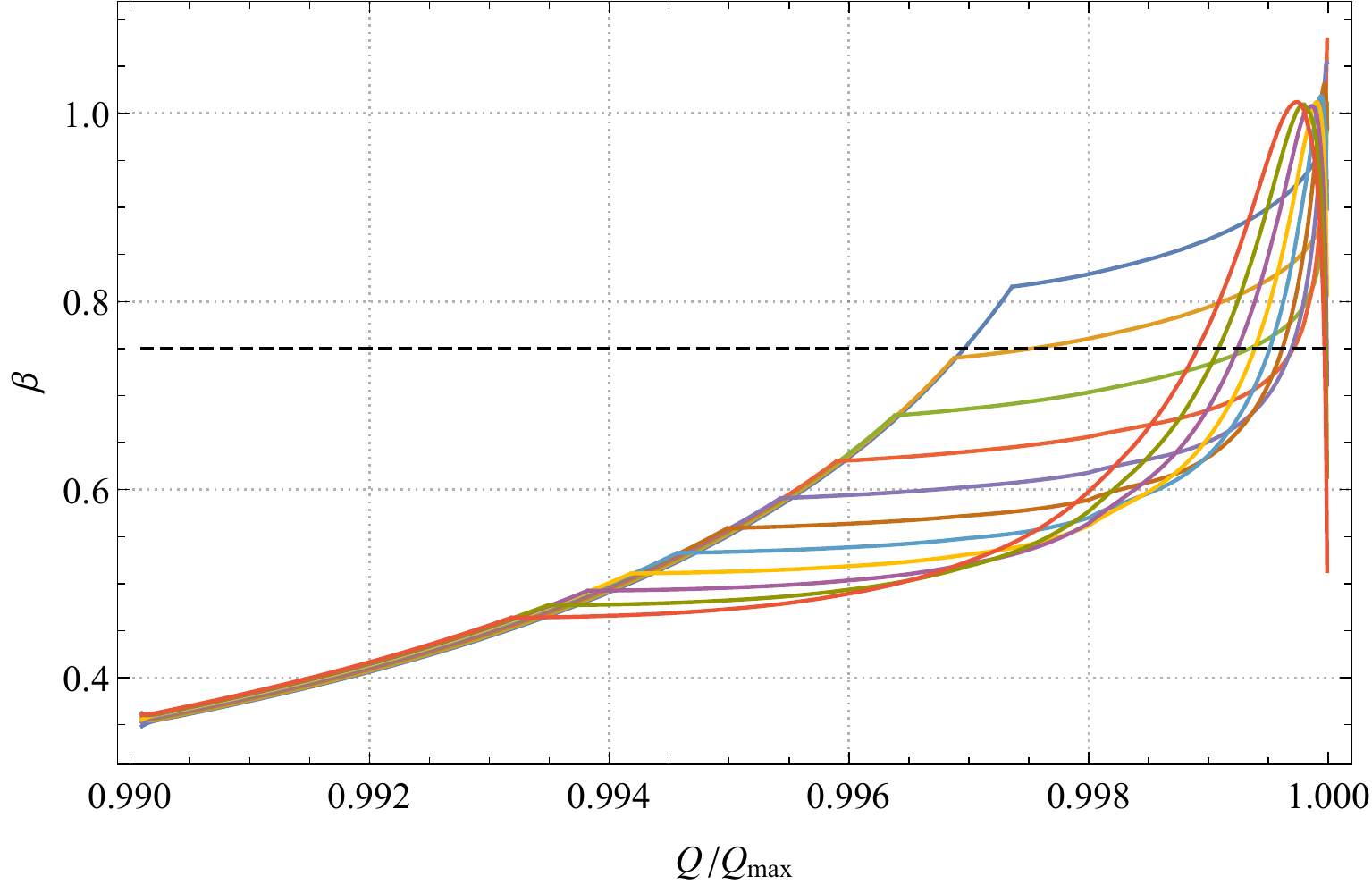}\label{a}
\end{minipage}
\begin{minipage}[b]{.18\linewidth}
\centering
\includegraphics[width=0.3in,height=1.4in]{fig/alpha.pdf}
\label{b}
\end{minipage}
\caption{Dominant modes $\beta=-\omega_I/\kappa_-$ for $\Lambda=0.14$ and $\alpha$ from $0$ to $0.2$, spaced $0.02$ apart.}
\label{fig32}
\end{figure}

Through the same process, we can plot the dominant modes for any $\alpha$. In this paper, we show the dominant mode of $\Lambda=0.06$ for different $\alpha$ in Fig. \ref{fig31} and the dominant mode of $\Lambda=0.14$ for different $\alpha$ in Fig. \ref{fig32}. The coupling constant is taken from $0$ to $0.2$, spaced $0.02$ apart, in these two figures. For the case where $\Lambda=0.06$, the curves of the dominant modes go down gradually as $\alpha$ increases, and there may appear an extreme point when the black hole is highly extremal. It is apparent that there does exist some $\alpha$ such that the dominant modes lie below the standard line $\beta=3/4$ for all nearly extremal black holes. The SCC will be recovered in these cases. However, when $\Lambda=0.14$, the dominant modes decrease as $\alpha$ increases only in a small interval. The maximum point also appears but they are all large than $3/4$. Therefore, for every $\alpha$ from $0$ to $0.2$, SCC is violated when $\Lambda=0.14$.

\begin{figure}
\centering
\includegraphics[width=0.5\textwidth]{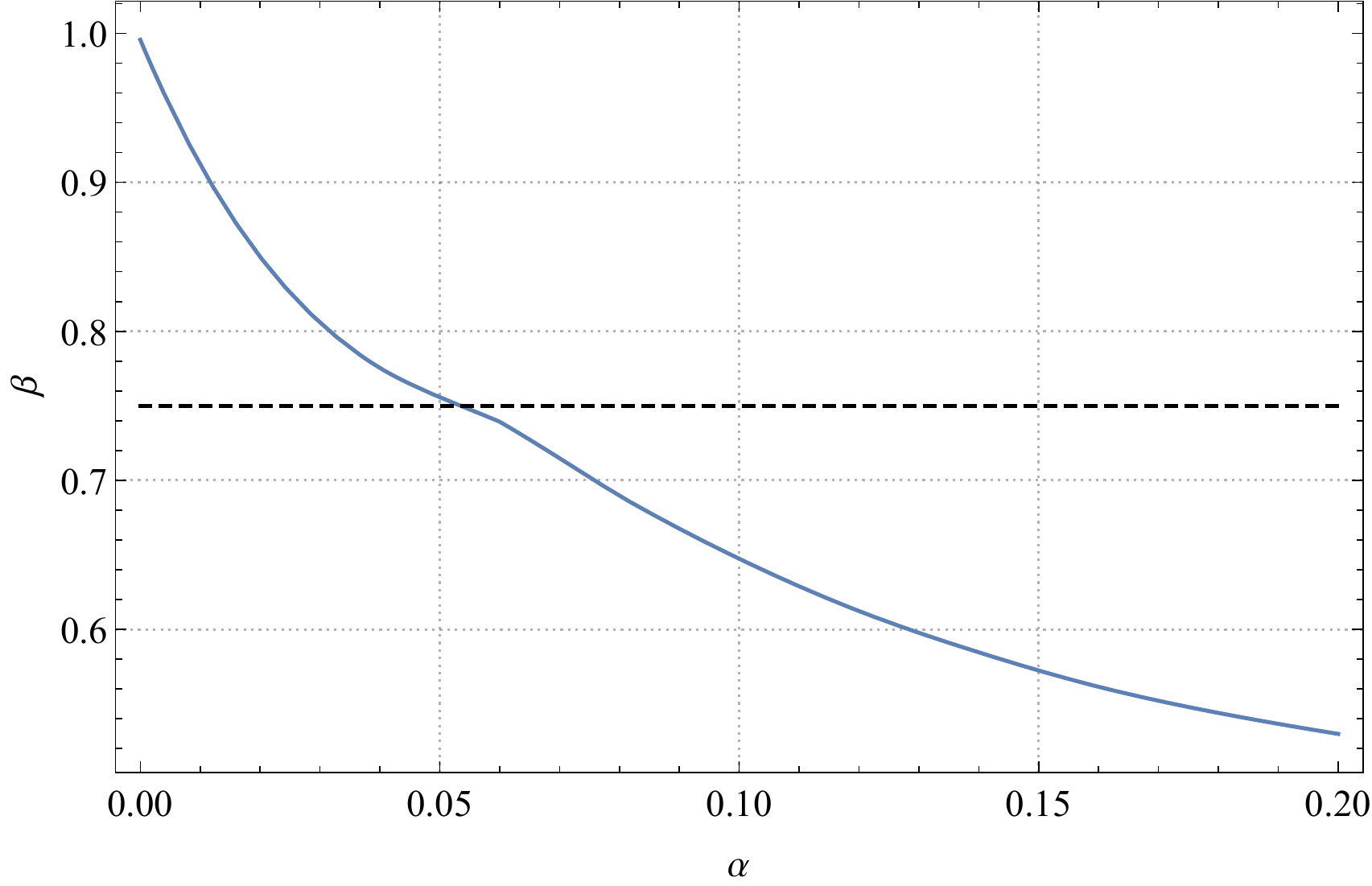}
\caption{Maximum values of $\beta=-{\omega_I}/{\kappa_-}$ for different $Q$ as a function of coupling constant $\alpha$.}
\label{fig4}
\end{figure}

Then, we would like to pay attention to the case of $\Lambda=0.06$, where the SCC might be recovered. We find the maximum point for each $\alpha$ and draw a curve of the maximum value varying with $\alpha$ in Fig. \ref{fig4}. We can find the maximum value is smaller than $3/4$ when $\alpha$ belongs to $(0.054,\,0.2)$. The SCC will be respected in this interval. In conclusion, the SCC can be recovered for some $\Lambda$ in the EMSGB theory, the range of $\alpha$ that can recover the SCC depends on the value of $\Lambda$.

\section{conclusion}\label{sec5}

In this paper, we investigate the SCC of the EMSGB theory. First, we introduce the gravitational theory and the geometry of the background spacetime. By considering the non-minimal coupled scalar field as a perturbation, we give the explicit expression of the equation of motion to describe the behavior of the scalar field. Next, through analyzing the extendibility of the solution, we find that the SCC will be violated when any $\beta=-\omega_I/\kappa_-$ is larger than $3/4$ where $\omega_I$ is the imaginary part of the QNM frequencies. Then, we calculate three families of QNM frequencies numerically for the nearly extremal RNdS black hole using the pseudospectral method and check the results by the direct integration method. The dominant modes which determine the fate of the SCC are extracted from these modes.

As a result, in EMSGB theory, we can find that the SCC can hardly be violated as usual when the black hole is not nearly extremal. However, for the near-extremal case, things are different as before. With the introduction of the non-minimal coupled scalar field, the SCC can be recovered by the non-minimal coupled scalar field for some appropriate cosmological constant $\Lambda$ and coupling constant $\alpha$. One can find the range of $\alpha$ such that the SCC is valid once a suitable $\Lambda$ is selected, the fate of SCC is determined by $\Lambda$ and $\alpha$ together.

\section*{Acknowledgement}

This research is supported by the Talents Introduction
Foundation of Beijing Normal University with Grant no. 310432102, and National Natural Science Foundation of China (NSFC) with Grants Nos. 11775022 and 11873044. We would like to thank Zhen Zhong for useful discussion.

\end{document}